\documentclass[aps,prd,twocolumn,floatfix,altaffilletter,superscriptaddress,preprintnumbers,
tightenlines,showpacs,showkeys,nofootinbib,notoccite]{revtex4-1}
\usepackage[utf8]{inputenc}
\usepackage[colorlinks=true,citecolor=blue,linkcolor=blue]{hyperref}
\usepackage[normalem]{ulem}
\usepackage{amsmath,amssymb, mathrsfs,esvect}
\usepackage{epsfig}
\usepackage{graphicx}               
\usepackage{url}
\usepackage{color}
\usepackage{slashed}
\usepackage{multirow}
\usepackage{placeins}
\usepackage[dvipsnames]{xcolor}
\usepackage{epstopdf}
\usepackage{soul}
\usepackage{tikz}
\usepackage[capitalise, english]{cleveref}
\usepackage{siunitx}
\usepackage{xspace}
\usepackage{booktabs}
\usetikzlibrary{trees}
\usetikzlibrary{decorations.pathmorphing}
\usetikzlibrary{decorations.markings}

\newcommand\myshade{80}
\colorlet{mylinkcolor}{ForestGreen}
\colorlet{mycitecolor}{Red}
\colorlet{myurlcolor}{violet}

\hypersetup{
	linkcolor  = mylinkcolor!\myshade!black,
	citecolor  = mycitecolor!\myshade!black,
	urlcolor   = myurlcolor!\myshade!black,
	colorlinks = true
}

\definecolor{jblue}{RGB}{20,50,100}
\definecolor{npurple}{RGB} {153, 51, 204}
\definecolor{wred}{RGB}{217,0,56}
\definecolor{white}{RGB}{255,255,255}

\allowdisplaybreaks

\usepackage{tikz,xcolor,hyperref}

\definecolor{lime}{HTML}{A6CE39}
\DeclareRobustCommand{\orcidicon}{\hspace{-1mm}
	\begin{tikzpicture}
		\draw[lime, fill=lime] (0,0) 
		circle [radius=0.16] 
		node[white] {{\fontfamily{qag}\selectfont \tiny \,ID}};
		\draw[white, fill=white] (-0.0525,0.095) 
		circle [radius=0.007];
	\end{tikzpicture}
	\hspace{-3mm}
}

\foreach \x in {A, ..., Z}{\expandafter\xdef\csname orcid\x\endcsname{\noexpand\href{https://orcid.org/\csname orcidauthor\x\endcsname}
		{\noexpand\orcidicon}}
}


\begin{document}

\title{Celestial Objects as Strongly-Interacting Non-Annihilating Dark Matter Detectors}

\author{Anupam Ray\orcidA{}}
\email{anupam.ray@berkeley.edu}
\affiliation{Department of Physics, University of California Berkeley, Berkeley, California 94720, USA}
\affiliation{School of Physics and Astronomy, University of Minnesota, Minneapolis, MN 55455, USA}

\date{\today}


\begin{abstract}
Non-annihilating dark matter particles, owing to their interactions with ordinary baryonic matter, can efficiently accumulate inside celestial objects. For heavy mass, they gravitate toward the core of the celestial objects, thermalize in a small core region, and eventually form tiny black holes via core collapse, resulting destruction of the host objects.  We demonstrate that the existence of a variety of celestial objects provides stringent constraints on strongly-interacting heavy dark matter, a blind-spot for the terrestrial dark matter detectors as well as for the cosmological probes. Celestial objects with larger sizes and lower core temperatures, such as Jupiter, are the most optimal detectors to probe the strongly-interacting heavy non-annihilating dark matter.
\end{abstract}

\maketitle
\preprint{N3AS-23-001}
\section{Introduction}
The existence of dark matter (DM) has been firmly established through its gravitational interactions
with the ordinary baryonic matter~\cite{Aghanim:2018eyx}. However, its identity still remains a mystery. The ongoing terrestrial and cosmological searches are trying to pinpoint its mass and hypothesized non-gravitational interactions with the baryonic matter~\cite{Cooley:2022ufh,Boddy:2022knd}. While some DM parameters have been excluded by these searches, many well-motivated candidates are yet to be explored. Strongly-interacting heavy non-annihilating DM is one such regime that remains largely untested. 

In this work, we demonstrate that celestial objects are excellent laboratories to search for strongly-interacting heavy non-annihilating DM. More specifically, we point out that the continued existence of a variety of stellar objects provides stringent exclusions on non-annihilating DM interactions over a wide mass range. We mainly answer two basic questions: why celestial objects are superior to search for  strongly-interacting heavy non-annihilating DM as compared to the terrestrial detectors, and which celestial objects are the most optimal targets?

Traditional direct detection experiments are not well-suited for heavy DM searches. The flux of Galactic DM at terrestrial detectors falls off linearly with higher DM mass, and the constraints weaken accordingly. Whereas, because of the enormous sizes of astrophysical objects and their long lifetimes, the effective exposure time $(\sim M_{\odot}\,\rm{Gyr})$ is orders of magnitude larger than human-made direct detection experiments $(\sim \rm{kT}\,\rm{yr})$, naturally providing sensitivity to the tiny flux of heavy DM.

Strongly-interacting DM is yet another blind-spot for typical direct detection experiments. If DM particles interact too strongly with baryonic matter, they lose a significant fraction of their energies via interactions with the material in the atmosphere as well as in the Earth-cover above the underground detectors. As a consequence, they slow down significantly and can not deposit sufficient amounts of energy for detection. Whereas, stellar objects are ideal probes for strongly-interacting DM as almost all the DM particles that transit through the stellar objects get trapped, leading to a maximal sensitivity.

Accumulation of particle DM in celestial objects is a key astrophysical probe of non-gravitational interactions of DM with the ordinary baryonic matter. DM particles from the galactic halo, owing to their interactions with the stellar constituents, can down-scatter to energies below the local escape energy, and become gravitationally bound to the stellar objects~\cite{1985ApJ...296..679P,Gould:1987ju,Gould:1987ir}. These bound DM particles lose more energy via repeated scatterings with the stellar constituents and eventually thermalize inside the stellar volume. Such bound thermalized DM particles can become abundant inside the stellar volume if they have sufficiently strong interactions with the baryonic matter, and possess intriguing phenomenological signatures.

For non-annihilating DM, such bound DM particles gradually accumulate, and for  heavy DM masses, they settle in a small region around the stellar core. Because of their prodigious abundance in a tiny core volume, their number density within the stellar core becomes quite large, allowing dark core collapse and subsequent black hole (BH) formation. This nascent BH, if not too light, can rapidly swallow the host, and the existence of stellar objects provides stringent constraints on non-annihilating DM interactions.

This scenario has been extensively studied in the context of old neutron stars~\cite{Goldman:1989nd,Gould:1989gw,Bertone:2007ae,deLavallaz:2010wp,McDermott:2011jp,Kouvaris:2010jy,Kouvaris:2011fi,Bell:2013xk,Guver:2012ba,Bramante:2013hn,Bramante:2013nma,Kouvaris:2013kra,Bramante:2014zca,Garani:2018kkd,Kouvaris:2018wnh,Dasgupta:2020dik,Lin:2020zmm,Dasgupta:2020mqg,Takhistov:2020vxs,Garani:2021gvc,Steigerwald:2022pjo,Singh:2022wvw} primarily because of their enormously large baryonic density as well as high compactness. More specifically, neutron stars can capture a significant number of DM particles from the Galactic halo even for the low DM-nucleon scattering cross-sections as the single-collision capture rate scales linearly with the compactness $(\sim M/R)$. Quantitatively, for a solar mass neutron star, residing in the solar neighborhood, with a typical radius of 10\,km, the capture rate is $\mathcal{O}(10^5)$ times larger than the Sun for low DM interactions. Apart from that, because of their large baryonic density $(\sim M/R^3)$,  the accumulated DM particles thermalize in a  tiny region around the core, implying a huge core-density favorable for transmutation. This indicates neutron stars are the most optimal targets to probe weakly interacting heavy non-annihilating DM, and the existence of old neutron stars in the solar neighborhood excludes $m_{\chi}=10^6$\,GeV and $\sigma_{\chi n}=10^{-48}$\,cm$^2$, the leading constraints on non-annihilating DM interactions~\cite{McDermott:2011jp,Garani:2018kkd}.

However, in the optically thick (large DM-nucleon scattering cross-section) regime, non-compact objects such as the Sun, Jupiter, and Earth are more suitable detectors to probe non-annihilating DM interactions. This is simply because in the optically thick regime, almost all of the DM particles that transit through a stellar object get trapped, and therefore, the accumulation rate increases with the larger size. Since the non-compact objects possess much larger radii, the accumulation rate for non-compact objects becomes comparable or even larger than the neutron stars  in the strongly-interacting regime.  Quantitatively, for $m_{\chi} =10^6$\,GeV, and sufficiently high $\sigma_{\chi n}$, the transit (accumulation) rate for a typical neutron star, residing in the solar neighborhood, is $ 10^{19}\,\rm{s}^{-1}$, comparable to the Earth, and $10^{-2}\,(10^{-5})$ smaller than the Jupiter (Sun). This naturally motivates us to explore the potential of non-compact stellar objects as strongly-interacting non-annihilating DM detectors. We found that stellar objects with relatively large radii and low core-temperature (such as Jupiter), are the most optimal detectors. This is simply because the total number of accumulated DM particles increases with a larger radius, and the BH formation becomes easier with a lower core-temperature, implying the most favorable transmutation criterion. Prior works~\cite{Starkman:1990nj,Kurita:2015vga,Acevedo:2020gro}, more particularly Ref.~\cite{Acevedo:2020gro} has recently explored the transmutation scenario for the Earth and the Sun. We systematically revisit the issue to gain more insight on the constraints, and we show that stellar objects with larger sizes and small core temperatures, such as Jupiter, provide the leading constraints on strongly-interacting heavy non-annihilating DM (see also Refs.~\cite{Leane:2020wob,Leane:2021tjj} for probing annihilating DM interactions with Jupiter-like planets).

The rest of the paper is organized as follows. In Section~\ref{cap}, we discuss different stages of DM-induced collapse of the celestial objects. In Section~\ref{results}, we present our exclusion limits from the existence of several stellar objects, demonstrating that the constraints obtained in this analysis cover new parts of the DM parameter space and bridge the gap between the cosmological probes~\cite{Gluscevic:2017ywp,Boddy:2018wzy,DES:2020fxi,Nadler:2019zrb}, and the terrestrial detectors~\cite{XENON:2017vdw,CRESST:2017ues,CRESST:2019jnq,CDMS:2002moo,Kavanagh:2017cru}. Finally, we summarize and conclude in Section~\ref{summ}.

\section{Dark Matter induced Collapse of Stellar Objects}\label{cap}
\begin{figure*}[!t]
	\centering
	\includegraphics[width=0.95\textwidth]{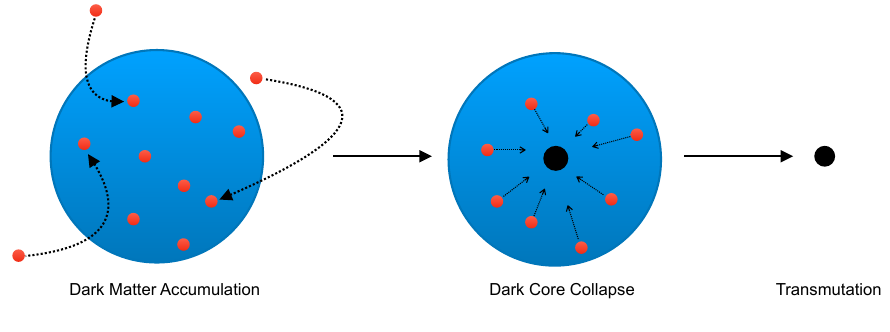}
	\caption{Transmutation of stellar objects via gradual accumulation of non-annihilating DM. Heavy DM  gravitate towards the core of the stellar objects and form tiny black holes via dark core collapse. These nascent BHs rapidly transmute the hosts by swallowing them, resulting in comparable low mass BHs.}
	\label{figure}
\end{figure*}

Non-annihilating DM particles can accumulate efficiently inside the stellar volume if they possess sufficiently strong interactions with the stellar nuclei. For heavy DM mass, they gravitate towards the stellar core and settle in a tiny core-region. Because of their prodigious abundance, and tiny core volume, their number density within the stellar core becomes tantalizingly larger, eventually resulting in a BH formation inside the stellar core.  This nascent BH, if not sufficiently light, can rapidly swallow the host, transmuting them to comparable mass BHs. In the following, we discuss different stages of DM-induced collapse of stellar objects, and a schematic diagram for this process is depicted in Fig.~\ref{figure}.

\subsection{Dark Matter Accumulation}
We first estimate the total number of captured DM particles inside the stellar volume. For clarity, we define the maximal capture rate as \textit{saturation capture rate} $(C_{\rm{sat}})$, and it occurs when all of the DM particles that transit through the stellar objects get trapped. For a particular velocity distribution of the incoming DM particles, the saturation capture rate is~\cite{Gould:1987ir}
\begin{equation}
C_{\rm{sat}} = \frac{\rho_{\chi}}{m_{\chi}} \pi R^2 \int \frac{f(u) du}{u} (u^2+v^2_{\rm{esc}})\,,
\end{equation}
where $\rho_{\chi} = 0.4$\,GeV/cm$^3$ is the Galactic DM density, $m_{\chi}$ is the DM mass, and $R$ is the radius of the stellar object. $f(u)$ denotes the velocity distribution of the incoming DM particles, with $v_{\rm{esc}}$ being the escape velocity of the stellar objects. For a Maxwell-Boltzmann velocity distribution, $C_{\rm{sat}}$ simplifies to
\begin{equation}
C_{\rm{sat}} = \frac{\rho_{\chi}}{m_{\chi}} \pi R^2 \sqrt{\frac{8}{3 \pi}} \bar{v} \left(1+\frac{3 v^2_{\rm{esc}}}{2 \bar{v}^2}\right)\,,
\end{equation}
where $\bar{v} = 270 $\,km/s denotes the average velocity of the DM particles in the Galactic halo.

A certain fraction of the DM particles will be captured by interacting with the stellar constituents, and we aim to estimate this capture fraction $(f_c)$. Since we are mostly interested in the optically thick regime (large DM-nucleon scattering cross-section), $f_c$ behaves differently for heavier and lighter DM. For heavier DM, i.e., when the DM mass $(m_{\chi})$ is larger than the target mass $(m_{A})$, scatterings do not alter the direction of the incoming DM particles. As a consequence, the trajectories of the incoming DM particles are not randomized, and they follow almost a linear trajectory that ends inside the stellar interior when the final velocity falls below the escape velocity. In this case, in the limit of multiple collisions, the DM particles are essentially guaranteed to be captured, resulting $f_c=1$~\cite{Bramante:2022pmn}. Whereas, in the opposite regime $(m_{\chi} < m_{A})$, the direction of the DM particles gets randomized after each collision, and as a result, a certain fraction of the DM particles can always escape from the stellar volume via reflection. This implies that for lighter DM, even for arbitrarily large cross-sections, the capture rate never reaches its saturation value $(f_c <1)$~\cite{Bramante:2022pmn}. 

For this analysis, we are interested in heavy DM capture inside stellar objects, and hence, we take $f_c=1$. It implies that in the optically thick regime, we take the capture rate to its saturation value, and it does not depend on the DM-nucleon scattering cross-section. For light DM capture in celestial objects, $f_c$ can be determined by the recent MCMC results~\cite{Bramante:2022pmn}, or from analytical estimates~\cite{Neufeld:2018slx}, both of which agree reasonably well.

In the optically thin regime (small DM-nucleon scattering cross-section), capture occurs via single scattering. This is simply because for smaller values of  DM-nucleon scattering cross-sections, the mean free path of the incoming DM particles becomes larger and becomes comparable to the size of the stellar objects, and as a result, they scatter once while transiting through the host. For this regime, we use the single-collision capture treatment~\cite{Gould:1987ir}, and  $f_c$ becomes substantially smaller.

\subsection{Spatial Distribution of Dark Matter inside Stellar Objects}
Captured DM particles rapidly thermalize inside the celestial objects for sufficiently high DM-nuclei scattering cross-sections~\cite{McDermott:2011jp,Kouvaris:2010jy,Kouvaris:2011fi,Bertoni:2013bsa,Garani:2018kkd,Acevedo:2020gro,Garani:2020wge}, and the spatial distribution of the thermalized DM particles inside the stellar volume depends on the effects of diffusion and gravity~\cite{Gould:1989hm,Banks:2021sba,Leane:2022hkk}. By considering the effects of diffusion and gravity in a self-consistent manner, the spatial distribution of the thermalized DM particles is~\cite{Leane:2022hkk} 
\begin{equation}\label{diffusion}
	\frac{\nabla n_{\chi}(r) }{n_{\chi}(r)}+ \left(\kappa +1 \right) 	\frac{\nabla T(r) }{T(r)}+\frac{m_{\chi} g(r)}{T(r)} = \frac{\Phi}{n_{\chi}(r) D_{\chi n}(r)} \frac{R_\oplus^2}{r^2}\,,
\end{equation}
where $n_{\chi}(r)$ denotes the number density of the thermalized DM particles within the stellar volume. $T(r)$ denotes the temperature profile of the celestial objects, and  $\Phi = C/\pi R_{\oplus}^2$ is the incoming flux of the DM particles, with $C$ denotes the capture rate.  $\kappa \sim -1/\left[2(1+m_{\chi}/m_{A})^{3/2}\right]$ and $D_{\chi n}\sim \lambda v_{\rm{th}}$ are the diffusion coefficients, where $\lambda$ denotes the mean free path of the DM particles and $v_{\rm{th}}$ denotes their thermal velocity. It is evident that, for very heavy DM mass, the diffusion co-efficient $(\kappa)$ becomes smaller as it scales as $m^{-3/2}_{\chi}$, and gravity (scales proportional to $m_{\chi}$) dominates over the diffusion processes. Therefore, heavy DM tends to settle down towards the core of the stellar objects. Quantitatively, by solving Eq.~(\ref{diffusion}) for  $n_{\chi}(r)$ with the boundary condition that the volume integral of
$n_{\chi}(r)$ provides the total number of captured DM particles, one can demonstrate that the captured DM particles mostly concentrate around the stellar core if they are heavy~\cite{Leane:2022hkk}.

Concentration of heavy DM around the stellar core can also be explained from the radius of the thermalization sphere $\left(r_{\rm{th}}=\sqrt{9k_BT/\left(4 \pi G \rho m _{\chi}\right)}\right)$, which results from the
balance between the thermal pressure and the gravitational potential~\cite{McDermott:2011jp,Kouvaris:2010jy}. Since, the radius of the thermalization sphere scales as $m^{-1/2}_{\chi}$, for heavy DM, $r_{\rm{th}}$ becomes smaller, indicating the concentration of the captured DM particles primarily occurs around the core.
\subsection{Dark Core Collapse \& Black Hole Formation}
For non-annihilating DM, accumulation grows linearly in time.  As a consequence, the number density of the captured DM particles inside the thermalization volume becomes tantalizingly large. Quantitatively, for DM mass of $10^6$ GeV, and sufficiently high DM-nuclei scattering cross-section (say $10^{-28}$\,cm$^2$), $\mathcal{O}(10^{36})$ number of DM particles thermalize inside the Earth within a radius of $\sim 6$ km, indicating a number density of $\sim 2 \times 10^{18}$\,cm$^{-3}$. This corresponds to a core density of  $\sim 2 \times 10^{24}$\,GeV cm$^{-3}$, around 25 orders of magnitude higher than the local Galactic DM density, and it further increases as $m^{3/2}_{\chi}$ for heavier DM masses. Once the core-density exceeds its critical threshold value, it undergoes a gravitational collapse, and eventually results in a BH formation inside the stellar core~\cite{Goldman:1989nd,Gould:1989gw,Bertone:2007ae,deLavallaz:2010wp,McDermott:2011jp,Kouvaris:2010jy,Kouvaris:2011fi,Bell:2013xk,Guver:2012ba,Bramante:2013hn,Bramante:2013nma,Kouvaris:2013kra,Bramante:2014zca,Garani:2018kkd,Kouvaris:2018wnh,Dasgupta:2020dik,Lin:2020zmm,Dasgupta:2020mqg,Acevedo:2020gro}. In the following, we quantify the critical core-density for the stellar objects.

The BH formation criterion via dark core collapse has been extensively discussed in the literature for compact stars~\cite{Goldman:1989nd,Gould:1989gw,Bertone:2007ae,deLavallaz:2010wp,McDermott:2011jp,Kouvaris:2010jy,Kouvaris:2011fi,Bell:2013xk,Guver:2012ba,Bramante:2013hn,Bramante:2013nma,Kouvaris:2013kra,Bramante:2014zca,Garani:2018kkd,Kouvaris:2018wnh,Dasgupta:2020dik,Lin:2020zmm,Dasgupta:2020mqg,Acevedo:2020gro,Garani:2021gvc,Steigerwald:2022pjo,Singh:2022wvw}, and is essentially determined by two conditions. The first one is within the stable thermal radius, the DM density has to exceed the corresponding baryonic density $(\rho_b)$. It leads to a self-gravitational collapse of the thermalized DM particles and is determined by~\cite{McDermott:2011jp,Kouvaris:2011fi,Garani:2018kkd}
\begin{equation}
\frac{m_{\chi} N^{\rm{self}}_{\chi}}{\frac{4}{3} \pi r^3_{\rm{th}}} \geq \rho_{b}\,,
\end{equation}
where $N^{\rm{self}}_{\chi}$ denotes the critical number of DM particles for ensuing self-gravitating collapse\footnote{This self-gravitating criterion is essentially equivalent to the Jeans instability criterion in Ref.~\cite{Acevedo:2020gro} up to $\mathcal{O}(1)$ factors.}. It  is independent of the spin of the DM particles and only depends on the DM mass as well as properties of the stellar objects such as core-density and core-temperature. For Earth, it corresponds to
\begin{equation}
 N^{\rm{self}}_{\chi} \sim 7\times 10^{36} \left(\frac{\rho_{\rm{core}}}{13\, \rm{g/cm^3}}\right) \left(\frac{T_{\rm{core}}}{5800\,\rm{K}}\right)^{3/2} \left(\frac{10^7\,  \rm{GeV}}{m_{\chi}}\right)^{5/2}\,, 
 \end{equation}
where, $\rho_{\rm{core}}$ denotes the core density, and $T_{\rm{core}}$ denotes the core temperature. The second condition is determined by the maximum number of DM particles that can be stabilized by the quantum degeneracy pressure and is commonly known as Chandrasekhar limit ($N^{\rm{cha}}_{\chi}$). Chandrasekhar limit depends on the spin-statistics of the DM particles as the quantum degeneracy pressure for bosonic DM stems from the Uncertainty principle, whereas, for fermionic DM, it arises from the Pauli exclusion principle. $N^{\rm{cha}}_{\chi}$ solely depends on the DM mass and for bosonic (fermionic) DM, it corresponds to $M^2_{\rm{pl}}/m^2_{\chi} \,(M^3_{\rm{pl}}/m^3_{\chi})$, where $M_{\rm{pl}}=1.22 \times 10^{19}$\,GeV denotes the Planck mass~\cite{McDermott:2011jp,Kouvaris:2011fi,Garani:2018kkd}. To summarize, for dark core collapse, the total number of captured DM particles inside a stellar object within its lifetime $(t_{\rm{age}})$ has to satisfy the following~\cite{McDermott:2011jp,Kouvaris:2011fi,Garani:2018kkd,Dasgupta:2020mqg}
\begin{equation}
N_{\chi} \rvert_{ t_{\rm{age}}} = C \times t_{\rm{age}} \geq \max \left[N^{\rm{self}}_{\chi}, N^{\rm{cha}}_{\chi}\right]\,.
\end{equation}
Note that, for stellar objects with high core temperature (such as Sun), the dark core collapse is essentially determined by $N^{\rm{self}}_{\chi}$ for bosonic as well as fermionic DM, leading to identical exclusion limits for bosonic/fermionic DM. Whereas, for stellar objects with low core temperature, the dark core collapse is determined by $N^{\rm{self}}_{\chi}\, (N^{\rm{cha}}_{\chi})$ for bosonic (fermionic) DM, leading to distinct exclusion limits for bosonic/fermionic DM.

\subsection{Growth and Evaporation of Nascent Black Holes}\label{aa}
It is important to stress that dark core collapse does not ensure the successful transmutation of the hosts. If the nascent BH is sufficiently light, transmutation can cease for two different reasons. Firstly, lighter BH takes a much longer time to swallow the hosts, and the swallow time $(\tau_{\rm{swallow}})$ can even be larger than the lifetime of the hosts. Secondly, and more importantly, Hawking radiation becomes significant for lighter BH masses ($\sim 1/M^2_{\rm{BH}}$), causing a rapid evaporation of the nascent BH. Since the mass of the nascent BH becomes smaller for heavier DM mass, this provides an upper limit on DM mass that can be probed via transmutation~\cite{McDermott:2011jp,Kouvaris:2011fi,Garani:2018kkd,Dasgupta:2020mqg}. We quantify the upper limits of $m_{\chi}$ for several stellar objects in the following, and it ranges around $\mathcal{O} (10^{10})$\,GeV for the stellar objects under consideration.

For the time-evolution of the nascent BH, we conservatively consider the baryonic matter accretion from the host (ignoring the possible DM accretion by the nascent BH)~\cite{McDermott:2011jp,Kouvaris:2011fi,Garani:2017jcj}
\begin{equation}
	\frac{dM_{\rm{BH}}}{dt} = \frac{4 \pi \rho_{\rm{core}} G^2M^2_{\rm{BH}}}{c^3_{s}} - \frac{P\,(M_{\rm{BH}})}{G^2M^2_{\rm{BH}}}\,,
\end{equation}
where $c_s=\sqrt{T_{\rm{core}}/m_n}$ denotes the  sound speed at the core of the stellar object, and $P\,(M_{\rm{BH}})$ denotes the Page factor~\cite{Page:1976df,MacGibbon:1991tj}. Page factor properly accounts into gray-body corrections of the Hawking evaporation spectrum, as well as the number of Standard Model (SM) species emission from an evaporating BH. In the classical black-body radiation limit, the Page factor evaluates to $1/\left(15360 \pi\right)$, and is commonly used in the literature.

Considering the gray-body corrections, the Page factor can be written as 
$2.8 \times 10^{-4} f(M_{\rm{BH}})$ where $f(M_{\rm{BH}})$ encodes the number of SM species emission from an evaporating BH~\cite{MacGibbon:1991tj}. For $M_{\rm{BH}} \geq 10^{17}$\,g (which only emit mass-less particles, such as photons and neutrinos), $f(M_{\rm{BH}})$ is normalized to unity~\cite{MacGibbon:1991tj}, and therefore, Page factor evaluates to $1/(1135\pi)$, an order of magnitude larger than the classical black-body limit. For $M_{\rm{BH}} \leq 10^{17}$\,g, the number of SM species emission from an evaporating BH crucially depends on its temperature (mass), and hence, $f(M_{\rm{BH}})$ varies with BH mass. We use the semi-analytic form of $f(M_{\rm{BH}})$ from Ref.~\cite{MacGibbon:1991tj} to estimate the Page factor in this regime. Quantitatively, for light BHs, i.e., $M_{\rm{BH}} \leq 10^{10}$\,g,  which can emit all SM species, $f(M_{\rm{BH}})$ evaluates to 15.35, and for $10^{15}\,\rm{g}\leq M_{\rm{BH}} \leq 10^{17}$\,g (which can emit electrons, positrons, photons, and neutrinos), $f(M_{\rm{BH}})$ evaluates to 1.569. To summarize, depending on BH mass, $f(M_{\rm{BH}})$ ranges from $\left(1 - 15.35\right)$, implying the range of Page factor from $1/(1135\pi)$ to $1/(74\pi)$. We verify that the Page factor obtained from the semi-analytic form of $f(M_{\rm{BH}})$ from Ref.~\cite{MacGibbon:1991tj} is in excellent agreement with the publicly available
\texttt{BlackHawk} package~\cite{Arbey:2021mbl}.

Since, the accretion term scales as $M^2_{\rm{BH}}$, and the evaporation term scales as $1/M^2_{\rm{BH}}$, for low BH masses, evaporation dominates over the accretion process. Quantitatively, for Sun, Jupiter, Earth, and Moon, Hawking evaporation dominates over the Bondi-Hoyle accretion for 
\begin{align}
	m_{\chi} \geq 
	\begin{cases} 
		7.1\times10^{11}\,\,\textrm{GeV}
		& 
		\\
		6.2\times10^{9}\,\,\textrm{GeV}
		& 
		\\
		2.4\,(5.3)\times10^{9}\,\,\textrm{GeV}
		& 
		\\
		1.0\,(6.0)\times10^{9}\,\,\textrm{GeV}
	\end{cases}
\end{align}
for non-annihilating bosonic (fermionic) DM. On the other hand, by requiring that the $\tau_{\rm{swallow}}$ has to be less than 1 Gyr, we obtain 
\begin{align}
	m_{\chi} \geq 
	\begin{cases} 
		2.1\times10^{10}\,\,\textrm{GeV}
		& 
		\\
		1.1\times10^{10}\,\,\textrm{GeV}
		& 
		\\
		0.9\,(1.5)\times10^{10}\,\,\textrm{GeV}
		& 
		\\
		0.8\,(3.1)\times10^{10}\,\,\textrm{GeV}
	\end{cases}
\end{align}
for non-annihilating bosonic (fermionic) DM. We note that, for stellar objects with  high core-temperature, $\tau_{\rm{swallow}}$ essentially determines the wash-out of the transmutation, whereas, for stellar objects with low core-temperature, it is determined by the efficient Hawking evaporation. This can simply be explained by the following. For high core-temperature stellar objects, the nascent BH becomes relatively larger $(M_{\rm{BH,init}} \sim T^{3/2})$, and therefore, the effects of Hawking evaporation become relatively sub-dominant, implying accretion determines the termination of the transmutation.

\subsection{Drift time and Maximal Possible Scattering Cross-section}
Transmutation of stellar objects also ceases at very high DM-nucleon scattering cross-sections. This is simply because, at very high DM-nucleon cross-sections, DM particles lose a significant amount of energy in the outer shells of these stellar objects, and might not reach the stellar core to form a micro BH. In other words, the viscous drag force that drives the DM particles toward the core results in a long drift time, and therefore, prohibits transmutation. We estimate the drift time by using the stellar density, temperature, and compositional profiles~\cite{Gould:1989gw,Bramante:2019fhi,Acevedo:2020gro}
\begin{equation}
	t_{\rm{drift}} = \frac{1}{G m_{\chi}} \sum_j \sigma_{\chi j} \int_{0}^{R} \frac{n_j(r) \sqrt{3A_jT(r)}}{\int_{0}^{r} d^3r^{\prime} \rho_j(r^{\prime})} dr\,,
\end{equation}
where $\sigma_{\chi j}$ denotes the DM-nuclei scattering cross-section, and it is related to the DM-nucleon scattering cross-section via $\sigma_{\chi j} = \sigma_{\chi n}\,A^2_j \left(\mu_{\chi A_j}/\mu_{\chi n}\right)^2$ with $A_j$ is the mass number of the $j$-th nuclei, and $\mu_{\chi n}$ is the reduced mass of the DM-nucleon system. We determine the ceilings of our results by demanding that $t_{\rm{drift}} \leq 1$\,Gyr. Quantitatively, for (Sun) Earth, it corresponds to
\begin{equation}
	\sigma_{\chi n} \leq (10^{-17})\, 10^{-21}\,\textrm{cm}^2 \,\left(\frac{m_{\chi}}{10^{7}\,\rm{GeV}}\right)\,,
\end{equation}
and is same for bosonic/fermionic DM.

\subsection{Properties of Stellar Objects}
We accurately estimate the capture rate and the transmutation criterion for these stellar objects by utilizing stellar object properties, such as density profiles, temperature profiles, and detailed chemical compositions. In the following, we provide the inputs that have been considered for this analysis. The density and temperature profiles for Sun, Earth, and Jupiter which have been used in this analysis have also complied in Ref.~\cite{Leane:2022hkk}.

\begin{itemize}

\item \textbf{Sun:}  We use the solar density and temperature profiles from~\cite{1996Sci...272.1286C}. For the chemical composition, we assume that the Sun is entirely made up of $\textsuperscript{1} \rm H$~\cite{Leane:2022hkk}.

\item \textbf{Jupiter:} We use the Jupiter density and temperature profiles from the Jovian model J11-4a~\cite{2012ApJS..202....5F}. For the chemical composition, we assume that Jupiter is entirely made up of $\textsuperscript{1} \rm H$~\cite{Leane:2022hkk}.

\item \textbf{Earth:} We use the Preliminary Reference Earth Model from~\cite{Dziewonski:1981xy} for the density profile, and we take the temperature profile from~\cite{article} under the assumption of a hydro-static equilibrium. For the chemical composition, we use the tabulated values from Ref.~\cite{Bramante:2019fhi} with the core-mantle boundary  at 3480\,km, and the mantle-crust boundary at 6346\,km. The core is dominantly made up of  $\textsuperscript{56} \rm Fe$, whereas, the mantle and the crust are dominantly made up of  $\textsuperscript{16} \rm O$.

\item \textbf{Moon:} We use the MAX model for density and the chemical compositions of the Moon~\cite{Garani:2019rcb}. We take the lunar core-mantle boundary at 450 km, and the mantle-crust boundary  at 1650 km. The lunar core is  dominantly made up of  $\textsuperscript{56} \rm Fe$, whereas, the mantle and the crust is dominantly made up of  $\textsuperscript{16} \rm O$. For the temperature profile, we consider the Moon as an isothermal sphere with $T=1700$\,K~\cite{Garani:2019rcb}.

\end{itemize}
It is important to stress that the interior modeling of the stellar objects, considered in this analysis, can have large uncertainties, and depending on these uncertainties, the exclusion regions can vary. For Moon and the Jupiter, the uncertainties are maximal~\cite{Garani:2019rcb,2012ApJ...750...52N,2016A&A...596A.114M}, and it primarily affects the self-gravitating criterion, $N^{\rm{self}}_{\chi}$. Since 
the transmutation criterion for bosonic DM is determined by the self-gravitating criterion ($N^{\rm{self}}_{\chi}$),  the exclusion regions for bosonic DM  differ based on these uncertainties. Whereas, the exclusion regions remain unaltered for fermionic DM as the transmutation criterion for fermionic DM is essentially determined by the Chandrasekhar limit ($N^{\rm{cha}}_{\chi}$), which is independent of the internal modeling of the celestial objects. Quantitatively, for the Moon, if we use MIN model~\cite{Garani:2019rcb}, we find that the exclusion region for bosonic DM weaken by a factor of $\sim 2$ as compared to the result obtained from MAX model.

\begin{figure*}[!t]	\includegraphics[width=0.4\textwidth]{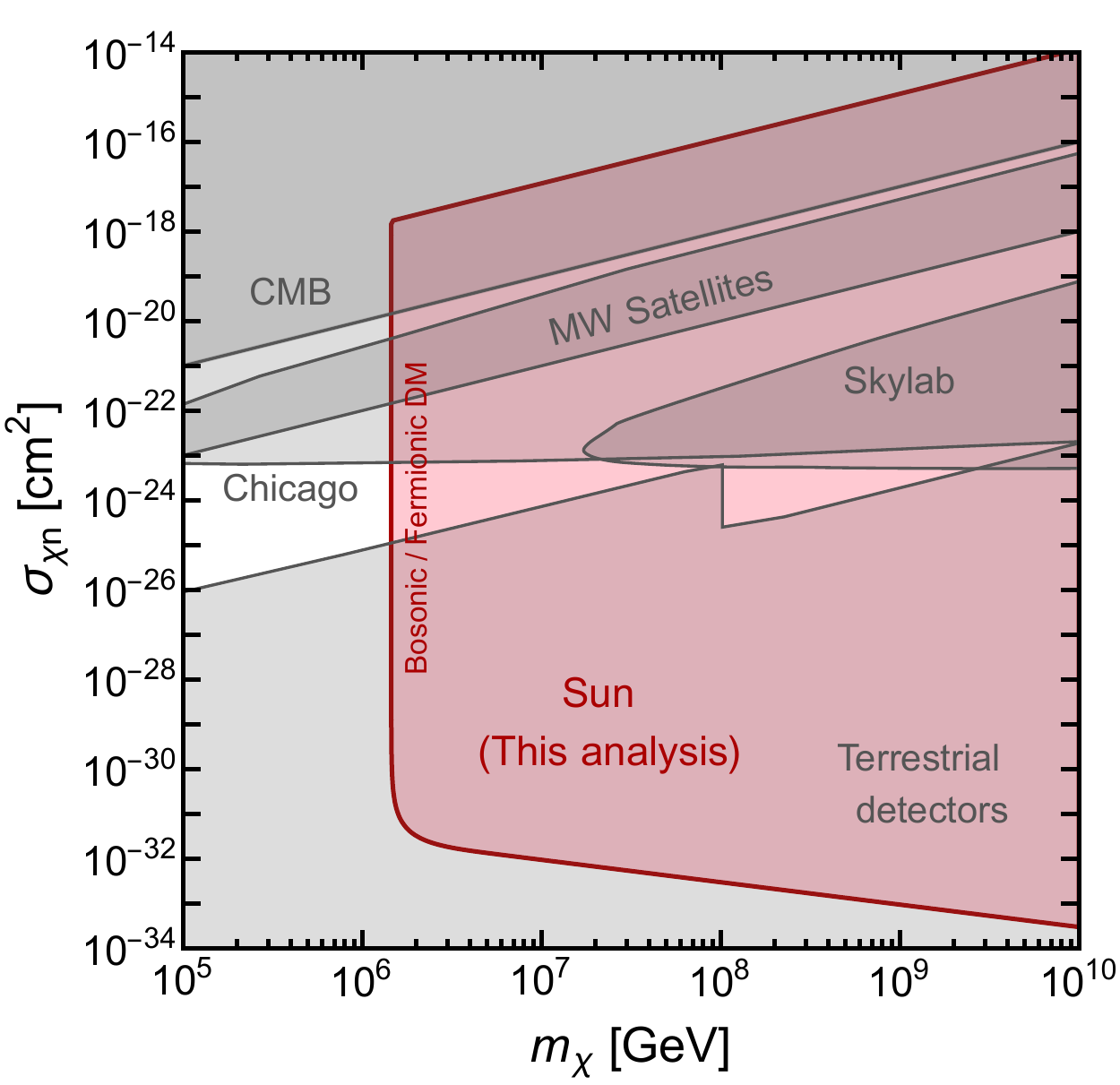}
	\hspace*{1.5 cm}
	\includegraphics[width=0.4\textwidth]{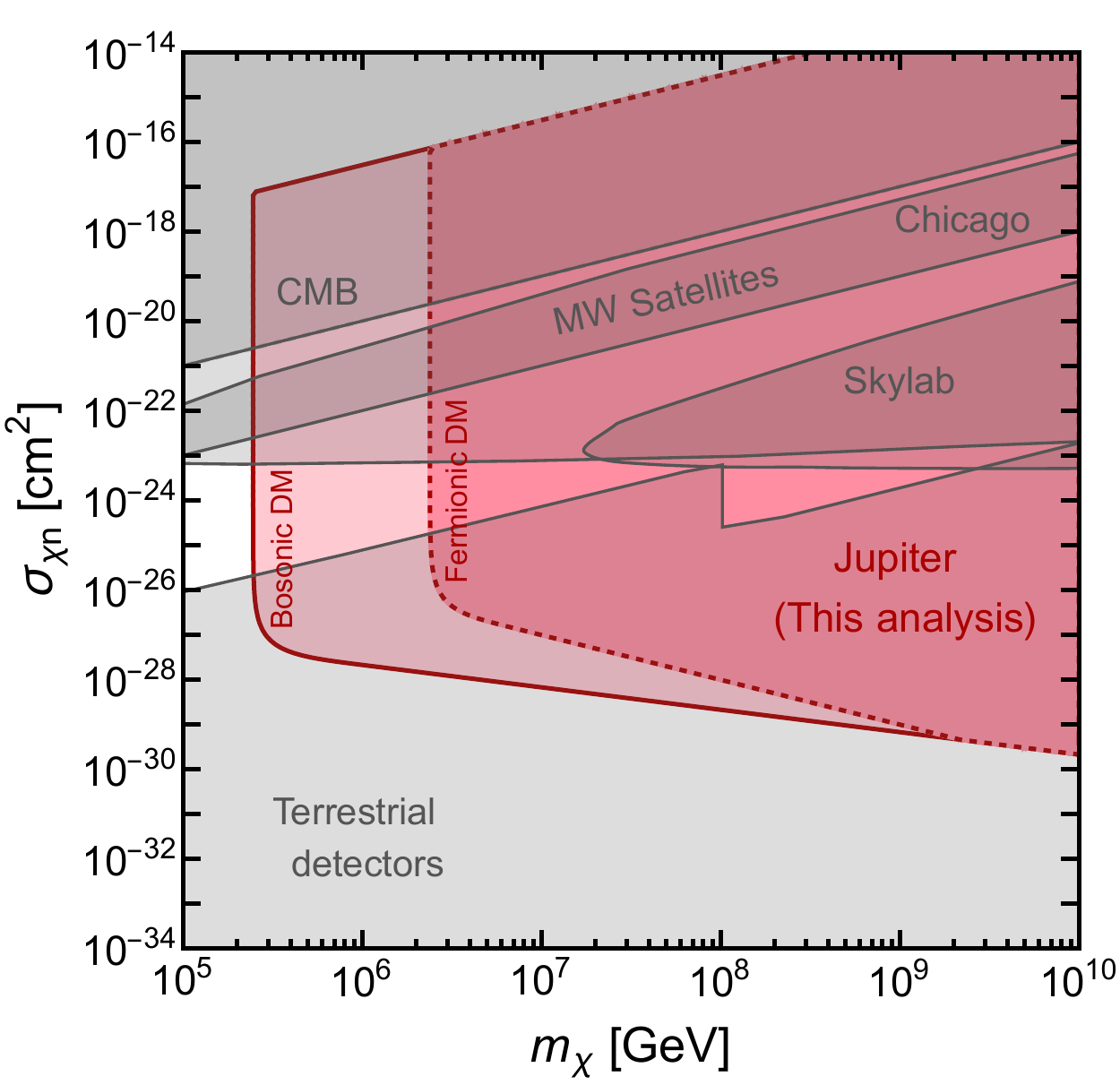}
	\includegraphics[width=0.4\textwidth]{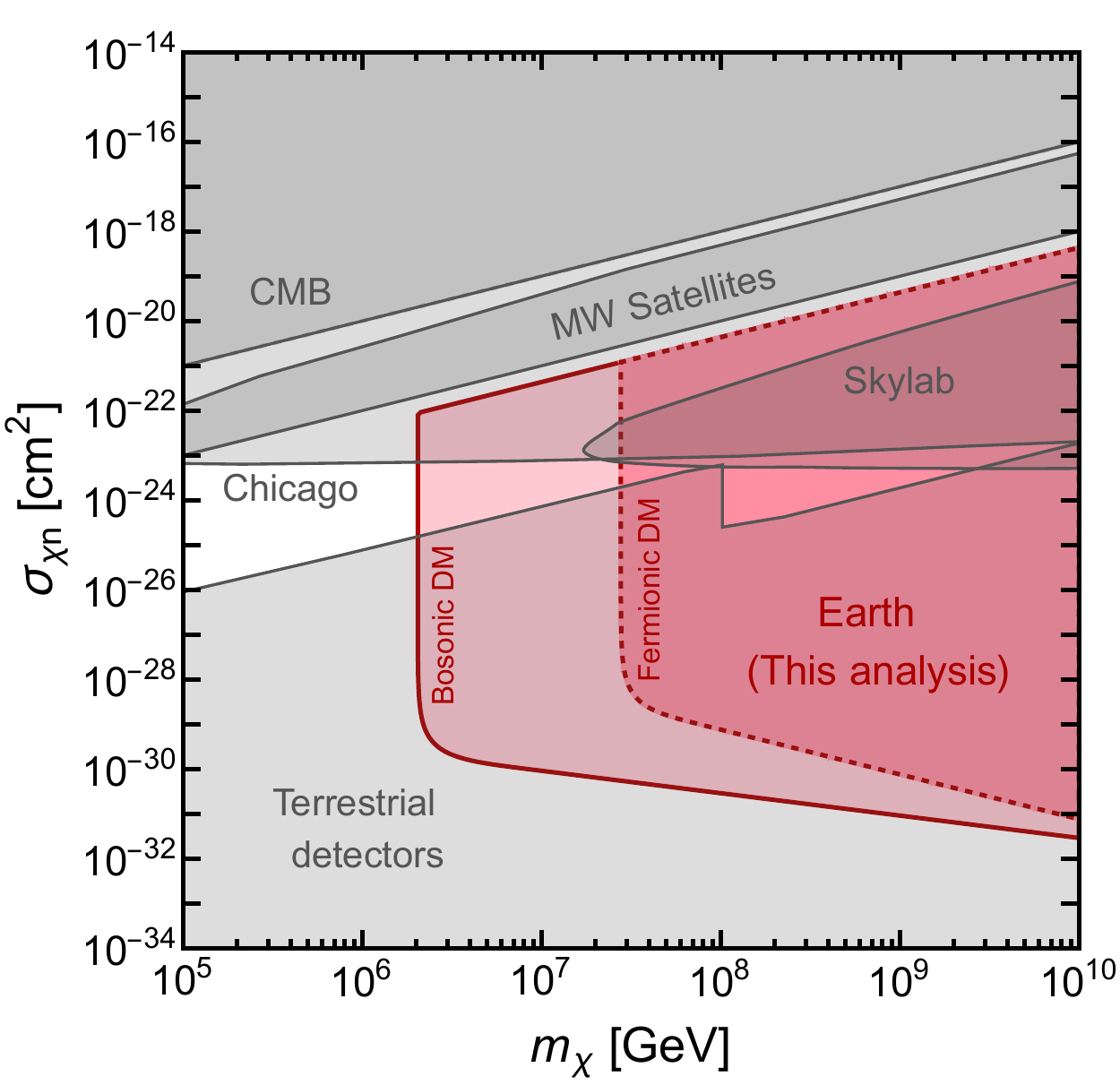}
	\hspace*{1.5 cm}
	\includegraphics[width=0.4\textwidth]{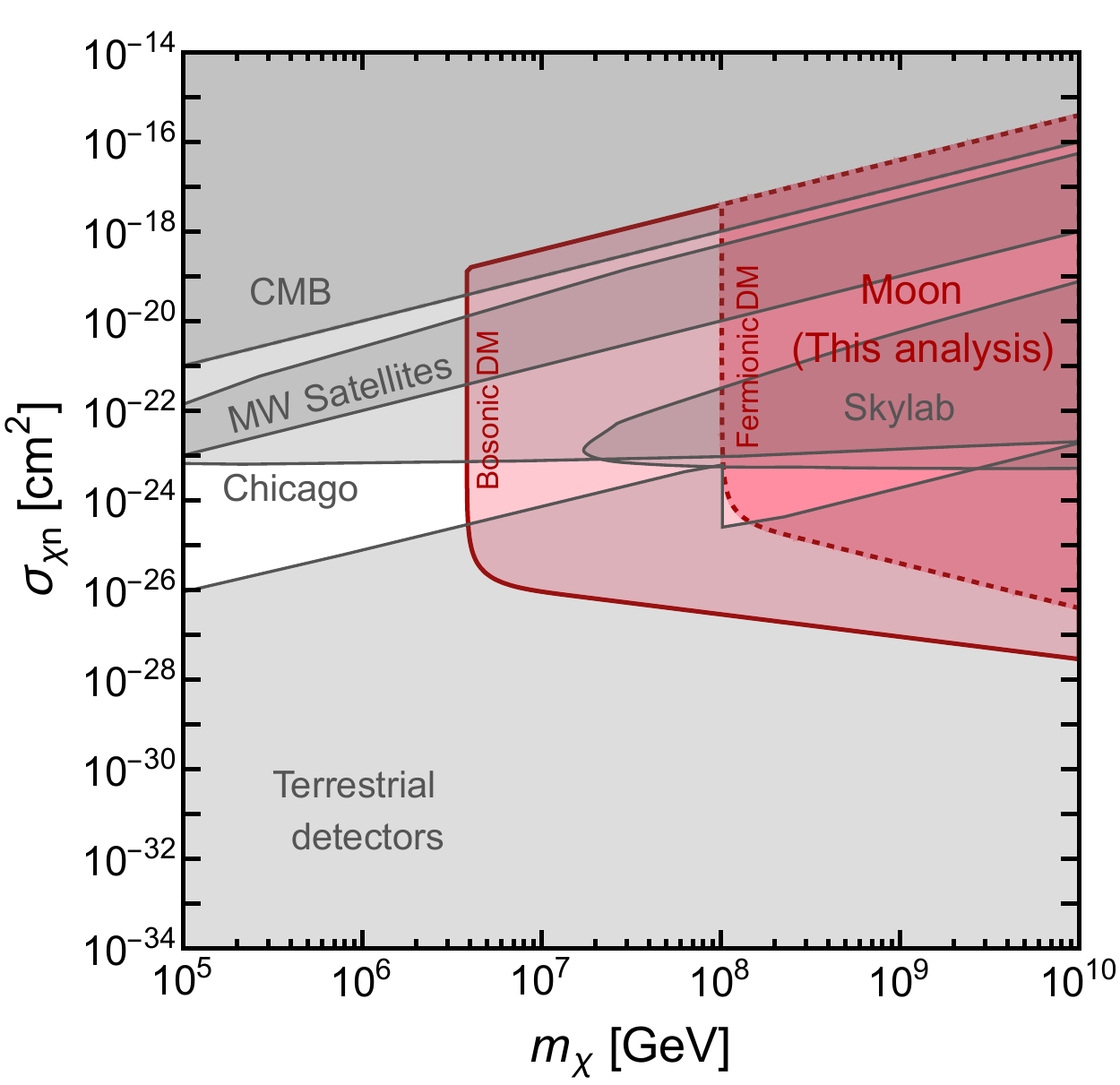}
	\caption{Exclusion limits on spin-independent DM-nucleon scattering cross-sections from the existence of several stellar objects. The regions shaded by the solid red lines correspond to non-annihilating bosonic DM, whereas, the regions shaded by dashed red lines correspond to non-annihilating fermionic DM. The top left (right) panel corresponds to the Sun (Jupiter), whereas, the bottom left (right) panel corresponds to the Earth (Moon). The existence of the stellar objects provides unprecedented sensitivity to  strongly-interacting heavy non-annihilating DM. We show the existing exclusion limits (gray shaded regions) from terrestrial searches~\cite{XENON:2017vdw,CRESST:2017ues,CRESST:2019jnq,CDMS:2002moo} (collected in~\cite{Kavanagh:2017cru,Digman:2019wdm,Carney:2022gse}), Skylab space station~\cite{Bhoonah:2020fys}, a recent shallow-depth experiment carried out at University of Chicago~\cite{Cappiello:2020lbk}, and cosmological measurements~\cite{Gluscevic:2017ywp,DES:2020fxi} for comparison. The existing constraints (gray-shaded regions) apply to both non-annihilating and  annihilating DM, whereas, the constraints obtained from this analysis (red-shaded regions) apply solely to non-annihilating/very feebly annihilating DM.}
	\label{exclusion}
\end{figure*}

\section{Results}\label{results}
We consider a variety of stellar objects, such as the Sun, Jupiter, Earth, and Moon, and demonstrate their potential as non-annihilating DM detectors. These choices are well-motivated by the fact they cover a wide range of size, density, and temperature, making them sensitive to different parts of the DM parameter space.  Sun has the largest size, and the highest core temperature, and as a consequence, the total number of captured DM particles as well as the threshold for transmutation both become higher. Jupiter has a somewhat larger size, but possesses a much lower core temperature, implying a higher capture rate but a lower threshold for transmutation. Earth and Moon have relatively smaller sizes and much lower core-temperatures, and hence, the capture rate as well as the threshold for transmutation both becomes smaller.

We show our main results in Fig.~\ref{exclusion} for non-annihilating bosonic and fermionic DM (spin-independent interactions). The top left (right) panel corresponds to Sun (Jupiter) as a DM detector, whereas, the bottom left (right) panel corresponds to Earth (Moon) as a DM detector. For stellar objects with low core temperatures, such as Moon, Earth, and Jupiter, the exclusion limit for bosonic DM is significantly stronger as compared to the fermionic DM. This is simply because for non-annihilating bosonic DM, the dark core collapse criterion is essentially determined by $N^{\rm{self}}_{\chi}$, whereas, for non-annihilating fermionic DM, it is determined by $N^{\rm{cha}}_{\chi}$, which is much higher than the  self-gravitating criterion, i.e, $N^{\rm{cha}}_{\chi,\,\rm{fermion}} \gg N^{\rm{self}}_{\chi}$. This implies that transmutation for low core-temperature stellar objects is much harder to attain for non-annihilating fermionic DM, explaining the weaker exclusions.  However, for stellar objects with higher core temperature, such as the Sun, the dark core collapse criterion for bosonic as well as fermionic DM is set by $N^{\rm{self}}_{\chi}$ (as it scales as $T^{3/2}_{\rm{core}}$), explaining identical exclusion limits for bosonic and fermionic DM.

The exclusion limits in Fig.~\ref{exclusion} can be  explained qualitatively from the following. For non-annihilating bosonic (fermionic) DM, the total number of captured DM particles linearly increases with lighter $m_{\chi}$, whereas, the threshold for transmutation increases as $m^{5/2}_{\chi}\,(m^3_{\chi})$, and hence, for light DM masses, transmutation can not be achieved. This explains the sharp vertical cut-offs in the low $m_{\chi}$ regime. For heavier DM masses, the mass of the nascent BH becomes smaller, resulting in two distinct effects. Firstly, the nascent BH takes a substantially longer time to consume the host, and secondly, Hawking evaporation becomes crucial. Because of these two effects, transmutation ceases, providing the vertical cut-offs around $m_{\chi} =10^{10}$\,GeV in Fig.~\ref{exclusion}. For the exact numerical values, see Section~\ref{aa}.

Next, we discuss the $\sigma_{\chi n}$ dependence of the exclusion limits in Fig.~\ref{exclusion}. For low DM-nucleon scattering cross-sections, the total number of captured DM particles within the stellar objects decreases, and eventually falls below the threshold for transmutation, indicating that low $\sigma_{\chi n}$ can not be probed via transmutation. Quantitatively, for non-annihilating bosonic DM, $\sigma_{\chi n} \leq 10^{-33}$\,cm$^2\,(\sigma_{\chi n} \leq 10^{-28}\,\rm{cm}^2)$ can not be probed by the existence of the Sun (Moon). Very high DM-nucleon scattering cross-sections are also inaccessible via transmutations. This is simply because for very high $\sigma_{\chi n}$, the drift time of the DM particles becomes much longer, and they can not reach the stellar core for large interactions. In Fig.~\ref{exclusion}, we show the ceilings of our results by demanding that $t_{\rm{drift}} \leq 1$\,Gyr. For the Sun, it corresponds to $\sigma_{\chi n} \leq 10^{-17}\,\textrm{cm}^2$ for $m_{\chi}=10^7$\,GeV, and linearly increases with heavier DM mass.

\subsection{Comparison with the Existing Constraints}
In Fig.~\ref{exclusion}, we show the existing constraints on spin-independent DM-nucleon scattering cross-section (gray-shaded regions) for comparison. The existing constraints  can be classified into three broad categories: astrophysical, cosmological, and terrestrial. Constraints obtained from the terrestrial direct detection experiments (labeled as Terrestrial detectors)~\cite{XENON:2017vdw,CRESST:2017ues,CRESST:2019jnq,CDMS:2002moo} are primarily based on non-observation of any anomalous scattering signature in the underground as well as in the surface detectors. We take these constraints from the summary plots in~\cite{Kavanagh:2017cru,Digman:2019wdm,Carney:2022gse}. Cosmological constraints, such as Planck measurements of temperature and polarization anisotropy of the cosmic microwave background (labeled as CMB)~\cite{Gluscevic:2017ywp,Boddy:2018wzy}, and Milky Way satellite observations (labeled as MW satellites)~\cite{DES:2020fxi,Nadler:2019zrb} are also shown for comparison.  Astrophysical constraints, such as disk stability~\cite{Starkman:1990nj}, interstellar gas cooling~\cite{Chivukula:1989cc}, and Galactic Center gas-cloud heating~\cite{Bhoonah:2018wmw,Bhoonah:2020dzs} are typically weaker than the constraints obtained from the Milky Way satellite observations, and therefore, are not shown for clarity. Very recently, a shallow-depth
experiment carried out at the University of Chicago (labeled as Chicago)~\cite{Cappiello:2020lbk} provides a stringent constraint on strongly-interacting heavy DM, and we show it in Fig.~\ref{exclusion}. MAJORANA demonstrator  at the Sanford underground
research facility~\cite{Clark:2020mna} and DEAP-3600 detector at SNOLAB~\cite{DEAPCollaboration:2021raj} also provide exclusions on strongly-interacting heavy DM  interactions $(m_{\chi} \gtrsim 10^8\,\rm{GeV})$. However, since these exclusions do not cover any additional parameter space in Fig.~\ref{exclusion}, as compared to the ``Terrestrial detectors"~\cite{Carney:2022gse}, they are not shown. Large panels of etched plastic, situated aboard the Skylab Space Station provide novel exclusion limits on DM-nucleon interactions (labeled as Skylab)~\cite{Starkman:1990nj,Wandelt:2000ad,Bhoonah:2020fys}. Rocket-based X-ray Quantum Calorimetry  (XQC) experiment~\cite{Erickcek:2007jv}, and searches for DM tracks in ancient underground mica~\cite{Price:1986ky,Bramante:2018tos} probe strongly-interacting heavy DM. However, the XQC limit and the mica limit primarily apply for $m_{\chi} \leq 10^5$\,GeV and $m_{\chi} \geq 10^{10}$\,GeV, respectively, and hence, are not shown. Finally, constraints obtained from cosmic ray silicon detector satellite (IMP7/8), and balloon-borne experiment (IMAX)~\cite{Wandelt:2000ad} are not based on detailed analyses in peer-reviewed papers, and therefore, are not shown in Fig.~\ref{exclusion}.

Comparing with the existing exclusion limits, it is evident that the existence of a variety of stellar objects provides novel constraints on strongly-interacting heavy non-annihilating DM. It provides unprecedented sensitivity to some regions in the parameter space as compared to the existing searches. Quantitatively, existence of the stellar objects exclude $\sigma_{\chi n}=10^{-24}$\,cm$^2$ for $m_{\chi}=10^7$\,GeV, not ruled by any other probes. This leading sensitivity simply stems from the fact that stellar objects, owing to their gigantic size and  long lifetime, can capture a copious amount of Galactic DM particles for sufficiently high $\sigma_{\chi n}$, eventually causing an implosion. We note that, stellar objects with relatively larger sizes with much lower core temperatures, such as Jupiter, are the optimal targets to probe transmutation. This is simply because the total number of accumulated DM particles quadratically increases with a larger radius, and the threshold for transmutation falls off with lower core-temperature, implying the most favorable transmutation criterion. It is also important to stress that  the existing constraints from cosmological and terrestrial searches apply for both non-annihilating and annihilating DM interactions, as they are solely based on scatterings. Whereas, the constraints obtained in this analysis apply only to non-annihilating/very-feebly annihilating DM particles.

We also note that, the exclusions obtained in Ref.~\cite{Acevedo:2020gro} for Earth are weaker than our analysis, however for Sun, the constraints are similar. This can be explained in the following. Ref.~\cite{Acevedo:2020gro} derived their results for bosonic DM with non-negligible (repulsive) quartic self-interactions. Because of the repulsive self-interactions among the DM particles, the effective Chandrasekhar limit substantially increases~\cite{Bell:2013xk}, and it essentially determines the dark core collapse criterion for Earth. Whereas, in our analysis, we consider non-annihilating bosonic DM, and the dark core collapse criterion for Earth is determined by $N^{\rm{self}}_{\chi}$, explaining the difference between the two analyses. For Sun, because of its much larger core-temperature, $N^{\rm{self}}_{\chi} (\sim T^{3/2}_{\rm{core}})$ always exceeds over the Chandrasekhar limit, and sets the dark core collapse criterion, resulting in a similar exclusion in both the analyses. For attractive self-interactions among the DM particles, bound-state formation occurs which affect the dynamics of dark core collapse~\cite{Bell:2013xk,Gresham:2018rqo,Garani:2022quc}.
\subsection{Anomalous Heating Signatures}
Non-annihilating DM can also heat up the stellar objects via successive scatterings with the stellar nuclei while getting captured. Such anomalous heating, commonly known as dark kinetic heating, can be observed in neutron stars with very low surface temperatures~\cite{Baryakhtar:2017dbj}. Cold neutron stars $(T_{\rm{NS}}=\mathcal{O}(1000)\,\rm{K})$ are ideal targets to search for dark kinetic heating as they have much higher escape velocities. Because of the large escape velocities, the velocity of the incoming DM particles gets significantly enhanced while falling into the steep gravitational potential of the neutron stars, and hence, they can transfer their kinetic energy to the stellar nuclei via collisions, heating up the cold neutron stars. We estimate the same for non-compact objects, concluding that because of their sufficiently low escape velocity, and larger size,  such anomalous heating signatures are too low to observe.

\section{Summary \& Conclusion}\label{summ}
Celestial objects, owing to their enormous size and long lifetime, naturally act as novel DM detectors. We demonstrate that the continued existence of the Sun, Jupiter, Earth, and Moon provides stringent constraints on  strongly-interacting non-annihilating DM interactions over a wide mass range.  These choices are well-motivated by the fact that each of these stellar objects (except the Moon, which  does not probe any additional parameter space as compared to the Jupiter and Sun, because of its much smaller size) is an optimal detector in a different regime as they cover a different range of size, density, and temperatures.  We probe regions in the DM parameter space, which are entirely inaccessible to the existing DM searches, demonstrating the novelty of our analysis. Our proposal is simply based on the fact that non-annihilating DM particles from the Galactic halo get captured inside the stellar objects if they interact sufficiently with the nucleons.  For heavy DM, these captured DM particles rapidly thermalize within a very small region around the stellar core, resulting in a tantalizingly large core-density. Once the core-density exceeds its critical threshold value, nascent BH forms inside the stellar core, eventually destroying the hosts. Mere existence of these stellar objects excludes such DM mass and cross-sections which predict the successful destruction of the hosts. We consider several stellar objects to demonstrate their potential as DM detectors and conclude that stellar objects with relatively larger sizes and low core temperatures, such as Jupiter, can probe the maximal DM mass window. Heavier DM masses are the most optimal for transmutation. However, as the nascent BH becomes smaller with an increase in DM mass, accretion becomes inefficient, and the Hawking evaporation becomes crucial, ceasing the transmutation  for heavy DM masses. Overall, given these stringent exclusion limits, our work naturally inspires similar analysis for other celestial objects, such as Brown Dwarfs, Red Giants, and Exoplanets,  once we have a better understanding of their density, temperature, and compositional properties. Finally, we point out an intriguing direction that the transmutation of celestial objects can lead to planetary mass BHs, possibly explaining the six ultrashort microlensing events observed in the OGLE data~\cite{Niikura:2019kqi}, NANOGrav detection of stochastic GW background~\cite{Domenech:2020ers}, as well as the BH hypothesis of Planet-9~\cite{Scholtz:2019csj}. Since, planet mass primordial BHs provide the viable solutions to this anomalies~\cite{Niikura:2019kqi,Domenech:2020ers,Scholtz:2019csj}, it would be interesting to explore the alternative solutions via transmutation.
\section{Acknowledgments} 
It is my pleasure to thank Basudeb
Dasgupta for helpful discussions  and useful comments on the manuscript. I also thank Javier F. Acevedo, Sulagna Bhattacharya, Nirmal Raj, Rebecca Leane, Maxim Pospelov, and Juri Smirnov for helpful exchanges. AR acknowledges support from the National Science Foundation (Grant No. PHY-2020275), and to the Heising-Simons Foundation (Grant No. 2017-228).

\bibliographystyle{JHEP}
\bibliography{ref.bib}

\providecommand{\href}[2]{#2}\begingroup\raggedright\begin{thebibliography}{10}

\bibitem{Aghanim:2018eyx}
{\scshape Planck} collaboration, N.~Aghanim et~al., \emph{{Planck 2018 results.
  VI. Cosmological parameters}},
  \href{https://doi.org/10.1051/0004-6361/201833910}{\emph{Astron. Astrophys.}
  {\bfseries 641} (2020) A6}
  [\href{https://arxiv.org/abs/1807.06209}{{\ttfamily 1807.06209}}].

\bibitem{Cooley:2022ufh}
J.~Cooley et~al., \emph{{Report of the Topical Group on Particle Dark Matter
  for Snowmass 2021}},  \href{https://arxiv.org/abs/2209.07426}{{\ttfamily
  2209.07426}}.

\bibitem{Boddy:2022knd}
K.~K. Boddy et~al., \emph{{Snowmass2021 theory frontier white paper:
  Astrophysical and cosmological probes of dark matter}},
  \href{https://doi.org/10.1016/j.jheap.2022.06.005}{\emph{JHEAp} {\bfseries
  35} (2022) 112} [\href{https://arxiv.org/abs/2203.06380}{{\ttfamily
  2203.06380}}].

\bibitem{1985ApJ...296..679P}
W.~H. {Press} and D.~N. {Spergel}, \emph{{Capture by the sun of a galactic
  population of weakly interacting, massive particles}},
  \href{https://doi.org/10.1086/163485}{\emph{Astrophysical Journal} {\bfseries
  296} (1985) 679}.

\bibitem{Gould:1987ju}
A.~Gould, \emph{{{WIMP} Distribution in and Evaporation From the Sun}},
  \href{https://doi.org/10.1086/165652}{\emph{Astrophys. J.} {\bfseries 321}
  (1987) 560}.

\bibitem{Gould:1987ir}
A.~Gould, \emph{{Resonant Enhancements in WIMP Capture by the Earth}},
  \href{https://doi.org/10.1086/165653}{\emph{Astrophys. J.} {\bfseries 321}
  (1987) 571}.

\bibitem{Goldman:1989nd}
I.~Goldman and S.~Nussinov, \emph{{Weakly Interacting Massive Particles and
  Neutron Stars}}, \href{https://doi.org/10.1103/PhysRevD.40.3221}{\emph{Phys.
  Rev. D} {\bfseries 40} (1989) 3221}.

\bibitem{Gould:1989gw}
A.~Gould, B.~T. Draine, R.~W. Romani and S.~Nussinov, \emph{{Neutron Stars:
  Graveyard of Charged Dark Matter}},
  \href{https://doi.org/10.1016/0370-2693(90)91745-W}{\emph{Phys. Lett. B}
  {\bfseries 238} (1990) 337}.

\bibitem{Bertone:2007ae}
G.~Bertone and M.~Fairbairn, \emph{{Compact Stars as Dark Matter Probes}},
  \href{https://doi.org/10.1103/PhysRevD.77.043515}{\emph{Phys. Rev. D}
  {\bfseries 77} (2008) 043515}
  [\href{https://arxiv.org/abs/0709.1485}{{\ttfamily 0709.1485}}].

\bibitem{deLavallaz:2010wp}
A.~de~Lavallaz and M.~Fairbairn, \emph{{Neutron Stars as Dark Matter Probes}},
  \href{https://doi.org/10.1103/PhysRevD.81.123521}{\emph{Phys. Rev. D}
  {\bfseries 81} (2010) 123521}
  [\href{https://arxiv.org/abs/1004.0629}{{\ttfamily 1004.0629}}].

\bibitem{McDermott:2011jp}
S.~D. McDermott, H.-B. Yu and K.~M. Zurek, \emph{{Constraints on Scalar
  Asymmetric Dark Matter from Black Hole Formation in Neutron Stars}},
  \href{https://doi.org/10.1103/PhysRevD.85.023519}{\emph{Phys. Rev. D}
  {\bfseries 85} (2012) 023519}
  [\href{https://arxiv.org/abs/1103.5472}{{\ttfamily 1103.5472}}].

\bibitem{Kouvaris:2010jy}
C.~Kouvaris and P.~Tinyakov, \emph{{Constraining Asymmetric Dark Matter through
  observations of compact stars}},
  \href{https://doi.org/10.1103/PhysRevD.83.083512}{\emph{Phys. Rev. D}
  {\bfseries 83} (2011) 083512}
  [\href{https://arxiv.org/abs/1012.2039}{{\ttfamily 1012.2039}}].

\bibitem{Kouvaris:2011fi}
C.~Kouvaris and P.~Tinyakov, \emph{{Excluding Light Asymmetric Bosonic Dark
  Matter}}, \href{https://doi.org/10.1103/PhysRevLett.107.091301}{\emph{Phys.
  Rev. Lett.} {\bfseries 107} (2011) 091301}
  [\href{https://arxiv.org/abs/1104.0382}{{\ttfamily 1104.0382}}].

\bibitem{Bell:2013xk}
N.~F. Bell, A.~Melatos and K.~Petraki, \emph{{Realistic neutron star
  constraints on bosonic asymmetric dark matter}},
  \href{https://doi.org/10.1103/PhysRevD.87.123507}{\emph{Phys. Rev. D}
  {\bfseries 87} (2013) 123507}
  [\href{https://arxiv.org/abs/1301.6811}{{\ttfamily 1301.6811}}].

\bibitem{Guver:2012ba}
T.~G\"uver, A.~E. Erkoca, M.~Hall~Reno and I.~Sarcevic, \emph{{On the capture
  of dark matter by neutron stars}},
  \href{https://doi.org/10.1088/1475-7516/2014/05/013}{\emph{JCAP} {\bfseries
  05} (2014) 013} [\href{https://arxiv.org/abs/1201.2400}{{\ttfamily
  1201.2400}}].

\bibitem{Bramante:2013hn}
J.~Bramante, K.~Fukushima and J.~Kumar, \emph{{Constraints on bosonic dark
  matter from observation of old neutron stars}},
  \href{https://doi.org/10.1103/PhysRevD.87.055012}{\emph{Phys. Rev. D}
  {\bfseries 87} (2013) 055012}
  [\href{https://arxiv.org/abs/1301.0036}{{\ttfamily 1301.0036}}].

\bibitem{Bramante:2013nma}
J.~Bramante, K.~Fukushima, J.~Kumar and E.~Stopnitzky, \emph{{Bounds on
  self-interacting fermion dark matter from observations of old neutron
  stars}}, \href{https://doi.org/10.1103/PhysRevD.89.015010}{\emph{Phys. Rev.
  D} {\bfseries 89} (2014) 015010}
  [\href{https://arxiv.org/abs/1310.3509}{{\ttfamily 1310.3509}}].

\bibitem{Kouvaris:2013kra}
C.~Kouvaris and P.~Tinyakov, \emph{{Growth of Black Holes in the interior of
  Rotating Neutron Stars}},
  \href{https://doi.org/10.1103/PhysRevD.90.043512}{\emph{Phys. Rev. D}
  {\bfseries 90} (2014) 043512}
  [\href{https://arxiv.org/abs/1312.3764}{{\ttfamily 1312.3764}}].

\bibitem{Bramante:2014zca}
J.~Bramante and T.~Linden, \emph{{Detecting Dark Matter with Imploding Pulsars
  in the Galactic Center}},
  \href{https://doi.org/10.1103/PhysRevLett.113.191301}{\emph{Phys. Rev. Lett.}
  {\bfseries 113} (2014) 191301}
  [\href{https://arxiv.org/abs/1405.1031}{{\ttfamily 1405.1031}}].

\bibitem{Garani:2018kkd}
R.~Garani, Y.~Genolini and T.~Hambye, \emph{{New Analysis of Neutron Star
  Constraints on Asymmetric Dark Matter}},
  \href{https://doi.org/10.1088/1475-7516/2019/05/035}{\emph{JCAP} {\bfseries
  05} (2019) 035} [\href{https://arxiv.org/abs/1812.08773}{{\ttfamily
  1812.08773}}].

\bibitem{Kouvaris:2018wnh}
C.~Kouvaris, P.~Tinyakov and M.~H.~G. Tytgat, \emph{{NonPrimordial Solar Mass
  Black Holes}},
  \href{https://doi.org/10.1103/PhysRevLett.121.221102}{\emph{Phys. Rev. Lett.}
  {\bfseries 121} (2018) 221102}
  [\href{https://arxiv.org/abs/1804.06740}{{\ttfamily 1804.06740}}].

\bibitem{Dasgupta:2020dik}
B.~Dasgupta, A.~Gupta and A.~Ray, \emph{{Dark matter capture in celestial
  objects: light mediators, self-interactions, and complementarity with direct
  detection}}, \href{https://doi.org/10.1088/1475-7516/2020/10/023}{\emph{JCAP}
  {\bfseries 10} (2020) 023}
  [\href{https://arxiv.org/abs/2006.10773}{{\ttfamily 2006.10773}}].

\bibitem{Lin:2020zmm}
G.-L. Lin and Y.-H. Lin, \emph{{Analysis on the black hole formations inside
  old neutron stars by isospin-violating dark matter with self-interaction}},
  \href{https://doi.org/10.1088/1475-7516/2020/08/022}{\emph{JCAP} {\bfseries
  08} (2020) 022} [\href{https://arxiv.org/abs/2004.05312}{{\ttfamily
  2004.05312}}].

\bibitem{Dasgupta:2020mqg}
B.~Dasgupta, R.~Laha and A.~Ray, \emph{{Low Mass Black Holes from Dark Core
  Collapse}}, \href{https://doi.org/10.1103/PhysRevLett.126.141105}{\emph{Phys.
  Rev. Lett.} {\bfseries 126} (2021) 141105}
  [\href{https://arxiv.org/abs/2009.01825}{{\ttfamily 2009.01825}}].

\bibitem{Takhistov:2020vxs}
V.~Takhistov, G.~M. Fuller and A.~Kusenko, \emph{{Test for the Origin of Solar
  Mass Black Holes}},
  \href{https://doi.org/10.1103/PhysRevLett.126.071101}{\emph{Phys. Rev. Lett.}
  {\bfseries 126} (2021) 071101}
  [\href{https://arxiv.org/abs/2008.12780}{{\ttfamily 2008.12780}}].

\bibitem{Garani:2021gvc}
R.~Garani, D.~Levkov and P.~Tinyakov, \emph{{Solar mass black holes from
  neutron stars and bosonic dark matter}},
  \href{https://doi.org/10.1103/PhysRevD.105.063019}{\emph{Phys. Rev. D}
  {\bfseries 105} (2022) 063019}
  [\href{https://arxiv.org/abs/2112.09716}{{\ttfamily 2112.09716}}].

\bibitem{Steigerwald:2022pjo}
H.~Steigerwald, V.~Marra and S.~Profumo, \emph{{Revisiting constraints on
  asymmetric dark matter from collapse in white dwarf stars}},
  \href{https://doi.org/10.1103/PhysRevD.105.083507}{\emph{Phys. Rev. D}
  {\bfseries 105} (2022) 083507}
  [\href{https://arxiv.org/abs/2203.09054}{{\ttfamily 2203.09054}}].

\bibitem{Singh:2022wvw}
D.~Singh, A.~Gupta, E.~Berti, S.~Reddy and B.~S. Sathyaprakash,
  \emph{{Constraining properties of asymmetric dark matter candidates from
  gravitational-wave observations}},
  \href{https://arxiv.org/abs/2210.15739}{{\ttfamily 2210.15739}}.

\bibitem{Starkman:1990nj}
G.~D. Starkman, A.~Gould, R.~Esmailzadeh and S.~Dimopoulos, \emph{{Opening the
  Window on Strongly Interacting Dark Matter}},
  \href{https://doi.org/10.1103/PhysRevD.41.3594}{\emph{Phys. Rev. D}
  {\bfseries 41} (1990) 3594}.

\bibitem{Kurita:2015vga}
Y.~Kurita and H.~Nakano, \emph{{Gravitational waves from dark matter collapse
  in a star}}, \href{https://doi.org/10.1103/PhysRevD.93.023508}{\emph{Phys.
  Rev. D} {\bfseries 93} (2016) 023508}
  [\href{https://arxiv.org/abs/1510.00893}{{\ttfamily 1510.00893}}].

\bibitem{Acevedo:2020gro}
J.~F. Acevedo, J.~Bramante, A.~Goodman, J.~Kopp and T.~Opferkuch, \emph{{Dark
  Matter, Destroyer of Worlds: Neutrino, Thermal, and Existential Signatures
  from Black Holes in the Sun and Earth}},
  \href{https://doi.org/10.1088/1475-7516/2021/04/026}{\emph{JCAP} {\bfseries
  04} (2021) 026} [\href{https://arxiv.org/abs/2012.09176}{{\ttfamily
  2012.09176}}].

\bibitem{Leane:2020wob}
R.~K. Leane and J.~Smirnov, \emph{{Exoplanets as Sub-GeV Dark Matter
  Detectors}},
  \href{https://doi.org/10.1103/PhysRevLett.126.161101}{\emph{Phys. Rev. Lett.}
  {\bfseries 126} (2021) 161101}
  [\href{https://arxiv.org/abs/2010.00015}{{\ttfamily 2010.00015}}].

\bibitem{Leane:2021tjj}
R.~K. Leane and T.~Linden, \emph{{First Analysis of Jupiter in Gamma Rays and a
  New Search for Dark Matter}},
  \href{https://arxiv.org/abs/2104.02068}{{\ttfamily 2104.02068}}.

\bibitem{Gluscevic:2017ywp}
V.~Gluscevic and K.~K. Boddy, \emph{{Constraints on Scattering of
  keV\textendash{}TeV Dark Matter with Protons in the Early Universe}},
  \href{https://doi.org/10.1103/PhysRevLett.121.081301}{\emph{Phys. Rev. Lett.}
  {\bfseries 121} (2018) 081301}
  [\href{https://arxiv.org/abs/1712.07133}{{\ttfamily 1712.07133}}].

\bibitem{Boddy:2018wzy}
K.~K. Boddy, V.~Gluscevic, V.~Poulin, E.~D. Kovetz, M.~Kamionkowski and
  R.~Barkana, \emph{{Critical assessment of CMB limits on dark matter-baryon
  scattering: New treatment of the relative bulk velocity}},
  \href{https://doi.org/10.1103/PhysRevD.98.123506}{\emph{Phys. Rev. D}
  {\bfseries 98} (2018) 123506}
  [\href{https://arxiv.org/abs/1808.00001}{{\ttfamily 1808.00001}}].

\bibitem{DES:2020fxi}
{\scshape DES} collaboration, E.~O. Nadler et~al., \emph{{Milky Way Satellite
  Census. III. Constraints on Dark Matter Properties from Observations of Milky
  Way Satellite Galaxies}},
  \href{https://doi.org/10.1103/PhysRevLett.126.091101}{\emph{Phys. Rev. Lett.}
  {\bfseries 126} (2021) 091101}
  [\href{https://arxiv.org/abs/2008.00022}{{\ttfamily 2008.00022}}].

\bibitem{Nadler:2019zrb}
E.~O. Nadler, V.~Gluscevic, K.~K. Boddy and R.~H. Wechsler, \emph{{Constraints
  on Dark Matter Microphysics from the Milky Way Satellite Population}},
  \href{https://doi.org/10.3847/2041-8213/ab1eb2}{\emph{Astrophys. J. Lett.}
  {\bfseries 878} (2019) 32}
  [\href{https://arxiv.org/abs/1904.10000}{{\ttfamily 1904.10000}}].

\bibitem{XENON:2017vdw}
{\scshape XENON} collaboration, E.~Aprile et~al., \emph{{First Dark Matter
  Search Results from the XENON1T Experiment}},
  \href{https://doi.org/10.1103/PhysRevLett.119.181301}{\emph{Phys. Rev. Lett.}
  {\bfseries 119} (2017) 181301}
  [\href{https://arxiv.org/abs/1705.06655}{{\ttfamily 1705.06655}}].

\bibitem{CRESST:2017ues}
{\scshape CRESST} collaboration, G.~Angloher et~al., \emph{{Results on
  MeV-scale dark matter from a gram-scale cryogenic calorimeter operated above
  ground}}, \href{https://doi.org/10.1140/epjc/s10052-017-5223-9}{\emph{Eur.
  Phys. J. C} {\bfseries 77} (2017) 637}
  [\href{https://arxiv.org/abs/1707.06749}{{\ttfamily 1707.06749}}].

\bibitem{CRESST:2019jnq}
{\scshape CRESST} collaboration, A.~H. Abdelhameed et~al., \emph{{First results
  from the CRESST-III low-mass dark matter program}},
  \href{https://doi.org/10.1103/PhysRevD.100.102002}{\emph{Phys. Rev. D}
  {\bfseries 100} (2019) 102002}
  [\href{https://arxiv.org/abs/1904.00498}{{\ttfamily 1904.00498}}].

\bibitem{CDMS:2002moo}
{\scshape CDMS} collaboration, D.~Abrams et~al., \emph{{Exclusion Limits on the
  WIMP Nucleon Cross-Section from the Cryogenic Dark Matter Search}},
  \href{https://doi.org/10.1103/PhysRevD.66.122003}{\emph{Phys. Rev. D}
  {\bfseries 66} (2002) 122003}
  [\href{https://arxiv.org/abs/astro-ph/0203500}{{\ttfamily
  astro-ph/0203500}}].

\bibitem{Kavanagh:2017cru}
B.~J. Kavanagh, \emph{{Earth scattering of superheavy dark matter: Updated
  constraints from detectors old and new}},
  \href{https://doi.org/10.1103/PhysRevD.97.123013}{\emph{Phys. Rev. D}
  {\bfseries 97} (2018) 123013}
  [\href{https://arxiv.org/abs/1712.04901}{{\ttfamily 1712.04901}}].

\bibitem{Bramante:2022pmn}
J.~Bramante, J.~Kumar, G.~Mohlabeng, N.~Raj and N.~Song, \emph{{Light Dark
  Matter Accumulating in Terrestrial Planets: Nuclear Scattering}},
  \href{https://arxiv.org/abs/2210.01812}{{\ttfamily 2210.01812}}.

\bibitem{Neufeld:2018slx}
D.~A. Neufeld, G.~R. Farrar and C.~F. McKee, \emph{{Dark Matter that Interacts
  with Baryons: Density Distribution within the Earth and New Constraints on
  the Interaction Cross-section}},
  \href{https://doi.org/10.3847/1538-4357/aad6a4}{\emph{Astrophys. J.}
  {\bfseries 866} (2018) 111}
  [\href{https://arxiv.org/abs/1805.08794}{{\ttfamily 1805.08794}}].

\bibitem{Bertoni:2013bsa}
B.~Bertoni, A.~E. Nelson and S.~Reddy, \emph{{Dark Matter Thermalization in
  Neutron Stars}},
  \href{https://doi.org/10.1103/PhysRevD.88.123505}{\emph{Phys. Rev. D}
  {\bfseries 88} (2013) 123505}
  [\href{https://arxiv.org/abs/1309.1721}{{\ttfamily 1309.1721}}].

\bibitem{Garani:2020wge}
R.~Garani, A.~Gupta and N.~Raj, \emph{{Observing the thermalization of dark
  matter in neutron stars}},
  \href{https://doi.org/10.1103/PhysRevD.103.043019}{\emph{Phys. Rev. D}
  {\bfseries 103} (2021) 043019}
  [\href{https://arxiv.org/abs/2009.10728}{{\ttfamily 2009.10728}}].

\bibitem{Gould:1989hm}
A.~Gould and G.~Raffelt, \emph{{Thermal Conduction of Massive Particles}},
  \href{https://doi.org/10.1086/168568}{\emph{Astrophys. J.} {\bfseries 352}
  (1990) 654}.

\bibitem{Banks:2021sba}
H.~Banks, S.~Ansari, A.~C. Vincent and P.~Scott, \emph{{Simulation of energy
  transport by dark matter scattering in stars}},
  \href{https://doi.org/10.1088/1475-7516/2022/04/002}{\emph{JCAP} {\bfseries
  04} (2022) 002} [\href{https://arxiv.org/abs/2111.06895}{{\ttfamily
  2111.06895}}].

\bibitem{Leane:2022hkk}
R.~K. Leane and J.~Smirnov, \emph{{Floating Dark Matter in Celestial Bodies}},
  \href{https://arxiv.org/abs/2209.09834}{{\ttfamily 2209.09834}}.

\bibitem{Garani:2017jcj}
R.~Garani and S.~Palomares-Ruiz, \emph{{Dark matter in the Sun: scattering off
  electrons vs nucleons}},
  \href{https://doi.org/10.1088/1475-7516/2017/05/007}{\emph{JCAP} {\bfseries
  05} (2017) 007} [\href{https://arxiv.org/abs/1702.02768}{{\ttfamily
  1702.02768}}].

\bibitem{Page:1976df}
D.~N. Page, \emph{{Particle Emission Rates from a Black Hole: Massless
  Particles from an Uncharged, Nonrotating Hole}},
  \href{https://doi.org/10.1103/PhysRevD.13.198}{\emph{Phys. Rev. D} {\bfseries
  13} (1976) 198}.

\bibitem{MacGibbon:1991tj}
J.~H. MacGibbon, \emph{{Quark and gluon jet emission from primordial black
  holes. 2. The Lifetime emission}},
  \href{https://doi.org/10.1103/PhysRevD.44.376}{\emph{Phys. Rev. D} {\bfseries
  44} (1991) 376}.

\bibitem{Arbey:2021mbl}
A.~Arbey and J.~Auffinger, \emph{{Physics Beyond the Standard Model with
  BlackHawk v2.0}},
  \href{https://doi.org/10.1140/epjc/s10052-021-09702-8}{\emph{Eur. Phys. J. C}
  {\bfseries 81} (2021) 910}
  [\href{https://arxiv.org/abs/2108.02737}{{\ttfamily 2108.02737}}].

\bibitem{Bramante:2019fhi}
J.~Bramante, A.~Buchanan, A.~Goodman and E.~Lodhi, \emph{{Terrestrial and
  Martian Heat Flow Limits on Dark Matter}},
  \href{https://doi.org/10.1103/PhysRevD.101.043001}{\emph{Phys. Rev. D}
  {\bfseries 101} (2020) 043001}
  [\href{https://arxiv.org/abs/1909.11683}{{\ttfamily 1909.11683}}].

\bibitem{1996Sci...272.1286C}
J.~{Christensen-Dalsgaard}, W.~{Dappen}, S.~V. {Ajukov}, E.~R. {Anderson},
  H.~M. {Antia}, S.~{Basu} et~al., \emph{{The Current State of Solar
  Modeling}},
  \href{https://doi.org/10.1126/science.272.5266.1286}{\emph{Science}
  {\bfseries 272} (1996) 1286}.

\bibitem{2012ApJS..202....5F}
M.~{French}, A.~{Becker}, W.~{Lorenzen}, N.~{Nettelmann}, M.~{Bethkenhagen},
  J.~{Wicht} et~al., \emph{{Ab Initio Simulations for Material Properties along
  the Jupiter Adiabat}},
  \href{https://doi.org/10.1088/0067-0049/202/1/5}{\emph{The Astrophysical
  Journal Supplement Series} {\bfseries 202} (2012) 5}.

\bibitem{Dziewonski:1981xy}
A.~M. Dziewonski and D.~L. Anderson, \emph{{Preliminary reference earth
  model}}, \href{https://doi.org/10.1016/0031-9201(81)90046-7}{\emph{Phys.
  Earth Planet. Interiors} {\bfseries 25} (1981) 297}.

\bibitem{article}
Y.~Zhang, J.-F. Lin, H.~He, F.~Liu, M.~Zhang, T.~Sato et~al., \emph{Shock
  compression and melting of an fe-ni-si alloy: Implications for the
  temperature profile of the earth's core and the heat flux across the
  core-mantle boundary},
  \href{https://doi.org/10.1002/2017JB014723}{\emph{Journal of Geophysical
  Research: Solid Earth} {\bfseries 123} (2018) }.

\bibitem{Garani:2019rcb}
R.~Garani and P.~Tinyakov, \emph{{Constraints on Dark Matter from the Moon}},
  \href{https://doi.org/10.1016/j.physletb.2020.135403}{\emph{Phys. Lett. B}
  {\bfseries 804} (2020) 135403}
  [\href{https://arxiv.org/abs/1912.00443}{{\ttfamily 1912.00443}}].

\bibitem{2012ApJ...750...52N}
N.~{Nettelmann}, A.~{Becker}, B.~{Holst} and R.~{Redmer}, \emph{{Jupiter Models
  with Improved Ab Initio Hydrogen Equation of State (H-REOS.2)}},
  \href{https://doi.org/10.1088/0004-637X/750/1/52}{\emph{The Astrophysical
  Journal} {\bfseries 750} (2012) 52}
  [\href{https://arxiv.org/abs/1109.5644}{{\ttfamily 1109.5644}}].

\bibitem{2016A&A...596A.114M}
Y.~{Miguel}, T.~{Guillot} and L.~{Fayon}, \emph{{Jupiter internal structure:
  the effect of different equations of state}},
  \href{https://doi.org/10.1051/0004-6361/201629732}{\emph{Astronomy \&
  Astrophysics} {\bfseries 596} (2016) A114}
  [\href{https://arxiv.org/abs/1609.05460}{{\ttfamily 1609.05460}}].

\bibitem{Digman:2019wdm}
M.~C. Digman, C.~V. Cappiello, J.~F. Beacom, C.~M. Hirata and A.~H.~G. Peter,
  \emph{{Not as big as a barn: Upper bounds on dark matter-nucleus cross
  sections}}, \href{https://doi.org/10.1103/PhysRevD.100.063013}{\emph{Phys.
  Rev. D} {\bfseries 100} (2019) 063013}
  [\href{https://arxiv.org/abs/1907.10618}{{\ttfamily 1907.10618}}].

\bibitem{Carney:2022gse}
D.~Carney et~al., \emph{{Snowmass2021 Cosmic Frontier White Paper: Ultraheavy
  particle dark matter}},  \href{https://arxiv.org/abs/2203.06508}{{\ttfamily
  2203.06508}}.

\bibitem{Bhoonah:2020fys}
A.~Bhoonah, J.~Bramante, B.~Courtman and N.~Song, \emph{{Etched plastic
  searches for dark matter}},
  \href{https://doi.org/10.1103/PhysRevD.103.103001}{\emph{Phys. Rev. D}
  {\bfseries 103} (2021) 103001}
  [\href{https://arxiv.org/abs/2012.13406}{{\ttfamily 2012.13406}}].

\bibitem{Cappiello:2020lbk}
C.~V. Cappiello, J.~I. Collar and J.~F. Beacom, \emph{{New experimental
  constraints in a new landscape for composite dark matter}},
  \href{https://doi.org/10.1103/PhysRevD.103.023019}{\emph{Phys. Rev. D}
  {\bfseries 103} (2021) 023019}
  [\href{https://arxiv.org/abs/2008.10646}{{\ttfamily 2008.10646}}].

\bibitem{Chivukula:1989cc}
R.~S. Chivukula, A.~G. Cohen, S.~Dimopoulos and T.~P. Walker, \emph{{Bounds on
  Halo Particle Interactions From Interstellar Calorimetry}},
  \href{https://doi.org/10.1103/PhysRevLett.65.957}{\emph{Phys. Rev. Lett.}
  {\bfseries 65} (1990) 957}.

\bibitem{Bhoonah:2018wmw}
A.~Bhoonah, J.~Bramante, F.~Elahi and S.~Schon, \emph{{Calorimetric Dark Matter
  Detection With Galactic Center Gas Clouds}},
  \href{https://doi.org/10.1103/PhysRevLett.121.131101}{\emph{Phys. Rev. Lett.}
  {\bfseries 121} (2018) 131101}
  [\href{https://arxiv.org/abs/1806.06857}{{\ttfamily 1806.06857}}].

\bibitem{Bhoonah:2020dzs}
A.~Bhoonah, J.~Bramante, S.~Schon and N.~Song, \emph{{Detecting composite dark
  matter with long-range and contact interactions in gas clouds}},
  \href{https://doi.org/10.1103/PhysRevD.103.123026}{\emph{Phys. Rev. D}
  {\bfseries 103} (2021) 123026}
  [\href{https://arxiv.org/abs/2010.07240}{{\ttfamily 2010.07240}}].

\bibitem{Clark:2020mna}
M.~Clark, A.~Depoian, B.~Elshimy, A.~Kopec, R.~F. Lang, S.~Li et~al.,
  \emph{{Direct Detection Limits on Heavy Dark Matter}},
  \href{https://doi.org/10.1103/PhysRevD.102.123026}{\emph{Phys. Rev. D}
  {\bfseries 102} (2020) 123026}
  [\href{https://arxiv.org/abs/2009.07909}{{\ttfamily 2009.07909}}].

\bibitem{DEAPCollaboration:2021raj}
{\scshape (DEAP Collaboration)\textdaggerdbl{}, DEAP} collaboration,
  P.~Adhikari et~al., \emph{{First Direct Detection Constraints on Planck-Scale
  Mass Dark Matter with Multiple-Scatter Signatures Using the DEAP-3600
  Detector}}, \href{https://doi.org/10.1103/PhysRevLett.128.011801}{\emph{Phys.
  Rev. Lett.} {\bfseries 128} (2022) 011801}
  [\href{https://arxiv.org/abs/2108.09405}{{\ttfamily 2108.09405}}].

\bibitem{Wandelt:2000ad}
B.~D. Wandelt, R.~Dave, G.~R. Farrar, P.~C. McGuire, D.~N. Spergel and P.~J.
  Steinhardt, \emph{{Selfinteracting dark matter}},  in \emph{{4th
  International Symposium on Sources and Detection of Dark Matter in the
  Universe (DM 2000)}}, pp.~263--274, 6, 2000,
  \href{https://arxiv.org/abs/astro-ph/0006344}{{\ttfamily astro-ph/0006344}}.

\bibitem{Erickcek:2007jv}
A.~L. Erickcek, P.~J. Steinhardt, D.~McCammon and P.~C. McGuire,
  \emph{{Constraints on the Interactions between Dark Matter and Baryons from
  the X-ray Quantum Calorimetry Experiment}},
  \href{https://doi.org/10.1103/PhysRevD.76.042007}{\emph{Phys. Rev. D}
  {\bfseries 76} (2007) 042007}
  [\href{https://arxiv.org/abs/0704.0794}{{\ttfamily 0704.0794}}].

\bibitem{Price:1986ky}
P.~B. Price and M.~H. Salamon, \emph{{Search for Supermassive Magnetic
  Monopoles Using Mica Crystals}},
  \href{https://doi.org/10.1103/PhysRevLett.56.1226}{\emph{Phys. Rev. Lett.}
  {\bfseries 56} (1986) 1226}.

\bibitem{Bramante:2018tos}
J.~Bramante, B.~Broerman, J.~Kumar, R.~F. Lang, M.~Pospelov and N.~Raj,
  \emph{{Foraging for dark matter in large volume liquid scintillator neutrino
  detectors with multiscatter events}},
  \href{https://doi.org/10.1103/PhysRevD.99.083010}{\emph{Phys. Rev. D}
  {\bfseries 99} (2019) 083010}
  [\href{https://arxiv.org/abs/1812.09325}{{\ttfamily 1812.09325}}].

\bibitem{Gresham:2018rqo}
M.~I. Gresham and K.~M. Zurek, \emph{{Asymmetric Dark Stars and Neutron Star
  Stability}}, \href{https://doi.org/10.1103/PhysRevD.99.083008}{\emph{Phys.
  Rev. D} {\bfseries 99} (2019) 083008}
  [\href{https://arxiv.org/abs/1809.08254}{{\ttfamily 1809.08254}}].

\bibitem{Garani:2022quc}
R.~Garani, M.~H.~G. Tytgat and J.~Vandecasteele, \emph{{Condensed dark matter
  with a Yukawa interaction}},
  \href{https://doi.org/10.1103/PhysRevD.106.116003}{\emph{Phys. Rev. D}
  {\bfseries 106} (2022) 116003}
  [\href{https://arxiv.org/abs/2207.06928}{{\ttfamily 2207.06928}}].

\bibitem{Baryakhtar:2017dbj}
M.~Baryakhtar, J.~Bramante, S.~W. Li, T.~Linden and N.~Raj, \emph{{Dark Kinetic
  Heating of Neutron Stars and An Infrared Window On WIMPs, SIMPs, and Pure
  Higgsinos}},
  \href{https://doi.org/10.1103/PhysRevLett.119.131801}{\emph{Phys. Rev. Lett.}
  {\bfseries 119} (2017) 131801}
  [\href{https://arxiv.org/abs/1704.01577}{{\ttfamily 1704.01577}}].

\bibitem{Niikura:2019kqi}
H.~Niikura, M.~Takada, S.~Yokoyama, T.~Sumi and S.~Masaki, \emph{{Constraints
  on Earth-mass primordial black holes from OGLE 5-year microlensing events}},
  \href{https://doi.org/10.1103/PhysRevD.99.083503}{\emph{Phys. Rev. D}
  {\bfseries 99} (2019) 083503}
  [\href{https://arxiv.org/abs/1901.07120}{{\ttfamily 1901.07120}}].

\bibitem{Domenech:2020ers}
G.~Dom\`enech and S.~Pi, \emph{{NANOGrav hints on planet-mass primordial black
  holes}}, \href{https://doi.org/10.1007/s11433-021-1839-6}{\emph{Sci. China
  Phys. Mech. Astron.} {\bfseries 65} (2022) 230411}
  [\href{https://arxiv.org/abs/2010.03976}{{\ttfamily 2010.03976}}].

\bibitem{Scholtz:2019csj}
J.~Scholtz and J.~Unwin, \emph{{What if Planet 9 is a Primordial Black Hole?}},
  \href{https://doi.org/10.1103/PhysRevLett.125.051103}{\emph{Phys. Rev. Lett.}
  {\bfseries 125} (2020) 051103}
  [\href{https://arxiv.org/abs/1909.11090}{{\ttfamily 1909.11090}}].

\end{thebibliography}\endgroup
\end{document}